%% file: root.tex
\documentclass[letterpaper, 10pt, conference]{ieeeconf}  

\input{commands.tex}

\IEEEoverridecommandlockouts                              

\title{\LARGE \bf
Blade Effective Wind Speed Estimation: A Subspace Predictive Repetitive Estimator Approach
}

\author{Yichao Liu$^{1}$, Atindriyo Kusumo Pamososuryo$^{1}$, Riccardo Ferrari$^{1}$, \\Tobias Gybel Hovgaard$^{2}$ and Jan-Willem van Wingerden$^{1}$
\thanks{This research was conducted in cooperation with Vestas Wind Systems A/S. It was also partially supported by the European Union via a Marie Sklodowska-Curie Action (Project EDOWE, grant 835901).
$^{1}$Delft University of Technology, Delft Center for Systems and Control, Mekelweg 2, 2628 CD Delft, The Netherlands.
        {\tt\small \{Y.Liu-17, A.K.Pamososuryo, R.Ferrari, J.W.vanWingerden\}@tudelft.nl}.
$^{2}$Vestas Technology R\&D, Denmark.
       {\tt\small togho@vestas.com}.}}

\begin{document}

\maketitle
\thispagestyle{empty}
\pagestyle{empty}

\begin{abstract}
\input{sections/0_abstract}
\end{abstract}

\section{Introduction}
\input{sections/1_introduction}

\section{Definition of the cone coefficient}\label{sec:2}
\input{sections/2_wind_turbine}

\section{Blade effective wind speed estimation}\label{sec:3}

\input{sections/3_SPRE.tex}

\section{Case study}\label{sec:4}
\input{sections/4_simulation}

\section{Conclusions}\label{sec:5}
\input{sections/5_conclusions}


\bibliographystyle{IEEEtran} 
\bibliography{references}


\end{document}

%% file: commands.tex
\usepackage{graphicx,keyval,trig,url} 
\usepackage[T1]{fontenc}     
\usepackage[utf8]{inputenc}	
\usepackage{graphicx}

\usepackage{amsthm}
\usepackage{amssymb}
\usepackage{amsmath}
\usepackage{mathtools}
\usepackage[mathscr]{eucal}
\usepackage{lipsum}
\usepackage{cuted} 
\usepackage{blindtext}
\usepackage{color}
\usepackage{dsfont}

\usepackage{epstopdf}
\usepackage{bm}   

\usepackage{color}
\usepackage{xcolor}

\newtheorem{rem}{Remark}
\usepackage{multirow}
\usepackage{longtable}

\usepackage{mathtools}
\usepackage[colorinlistoftodos]{todonotes}

%% file: sections/0_abstract.tex
Modern wind turbine control algorithms typically utilize rotor effective wind speed measured from an anemometer on the turbine's nacelle.
Unfortunately, the measured wind speed from such a single measurement point does not give a good representation of the effective wind speed over the blades, as it does not take the varying wind condition within the entire rotor area into account. As such, Blade Effective Wind Speed (BEWS) estimation can be seen as a more accurate alternative.
This paper introduces a novel Subspace Predictive Repetitive Estimator (SPRE) approach to estimate the BEWS using blade load measurements.
In detail, the azimuth-dependent cone coefficient is firstly formulated to describe the mapping between the out-of-plane blade root bending moment and the wind speed over blades. 
Then, the SPRE scheme, which is inspired by Subspace Predictive Repetitive Control (SPRC), is proposed to estimate the BEWS.
Case studies exhibit the proposed method's effectiveness at predicting BEWS and identifying wind shear in varying wind speed conditions.
Moreover, this novel technique enables complicated wind inflow conditions, where a rotor is impinged and overlapped by wake shed from an upstream turbine, to be estimated.



%% file: sections/1_introduction.tex
In the past several decades, wind energy has been playing an increasingly important role in the international energy mix with the global wind industry reached a milestone of 651\,GW cumulative installed capacity in 2019, with the rapid growth of 10\% compared to the previous year~\cite{GWEC_2020}. 
Modern wind turbines tend to have larger rotor diameters and more slender towers, which lead to an increase in dynamic loadings on the turbines~\cite{Veers_2019}.
This induces a growing demand for more advanced wind turbine control systems.

In designing an advanced wind turbine controller, the inaccuracy of effective wind speed information becomes one of the arising issues as the discrepancies between the actual and measured wind speeds might deteriorate the control performance to some extent~\cite{Soltani_2013}.
Having a better understanding of the wind inflow conditions on an operating wind turbine would be beneficial in developing more effective controllers.

Such detailed knowledge of the effective wind speed over the rotor area, however, is not directly available through conventional measurements.
An anemometer, which is typically mounted on a wind turbine's nacelle, can only measure wind speed and direction at a single point in space where the device is deployed and, thus, insufficient to observe the effective wind speed affecting the entire rotor.
Therefore, it becomes compelling to consider wind sensing methodologies as alternatives.

The notion of wind sensing, which utilizes wind turbine rotor as a generalized anemometer, has been proposed in the literature~\cite{Soltani_2013} and stems from the fact that any changes in the wind inflow conditions can be identified by the changes in the rotor response, such as blade loads or accelerations, blade pitch, torque, \emph{etc}.
Recent literature~\cite{Bottasso_2018}, implied that Rotor Effective Wind Speed (REWS) along with several wind states, \emph{e.g.,} wind shear, yaw misalignment, can be identified based on blade loads measurements.
Unfortunately, the current algorithms in load-sensing approaches are not able to account for the periodic wind flow and disturbance over an individual blade which could be proven useful for advanced control strategies~\cite{Simley2013}.
In addition, the capability of estimating the time-varying wind speed, wind shear and wake impingement is still questionable.

In this paper, we propose a novel estimator which takes the time-varying wind speed into account. 
This is achieved by assuming that the Blade Effective Wind Speed (BEWS) is periodic and added to the slowly varying Rotor Effective Wind Speed (REWS).
In detail, an azimuth-dependent cone coefficient is defined to capture the blade loads. 
With the aid of such coefficient, the BEWS can then be estimated based on a wind speed estimator called Subspace Predictive Repetitive Estimator (SPRE), which is inspired by the Subspace Predictive Repetitive Control (SPRC) approach.
The basic idea of SPRC was initially proposed by van Wingerden \textit{et al.}~\cite{Wingerden_2011}, which showed promising results in load-limiting individual pitch control~\cite{Navalkar_2014,Navalkar_2015, frederik2018} and fault-tolerant control~\cite{liu2020faulttolerant, liu2020acc, liu2020fast} of wind turbines.
Considering that the wind inflow condition over the rotor is periodic, a SPRE approach is then inspired and developed to estimate the effective wind speed over an individual blade.
In order to verify the developed SPRE approach, in this paper a series of case studies is performed on a 5MW wind turbine reference model developed by the U.S. National Renewable Energy Laboratory (NREL)~\cite{Jonkman_2009}.

The structure of this paper is organized as follows.
Section \ref{sec:2} outlines the wind turbine model and the simulation environment. 
In Section \ref{sec:3}, the methodology of the proposed wind speed estimator approach, including the definition of azimuth-dependent cone coefficient and the SPRE algorithm is elaborated. 
Subsequently, several case studies are carried out to demonstrate the effectiveness of the developed approach in Section \ref{sec:4}.
Conclusions are drawn from these case studies and discussed in Section \ref{sec:5}.

%% file: sections/2_wind_turbine.tex

\begin{figure}
\centering 
\includegraphics[width=1\columnwidth]{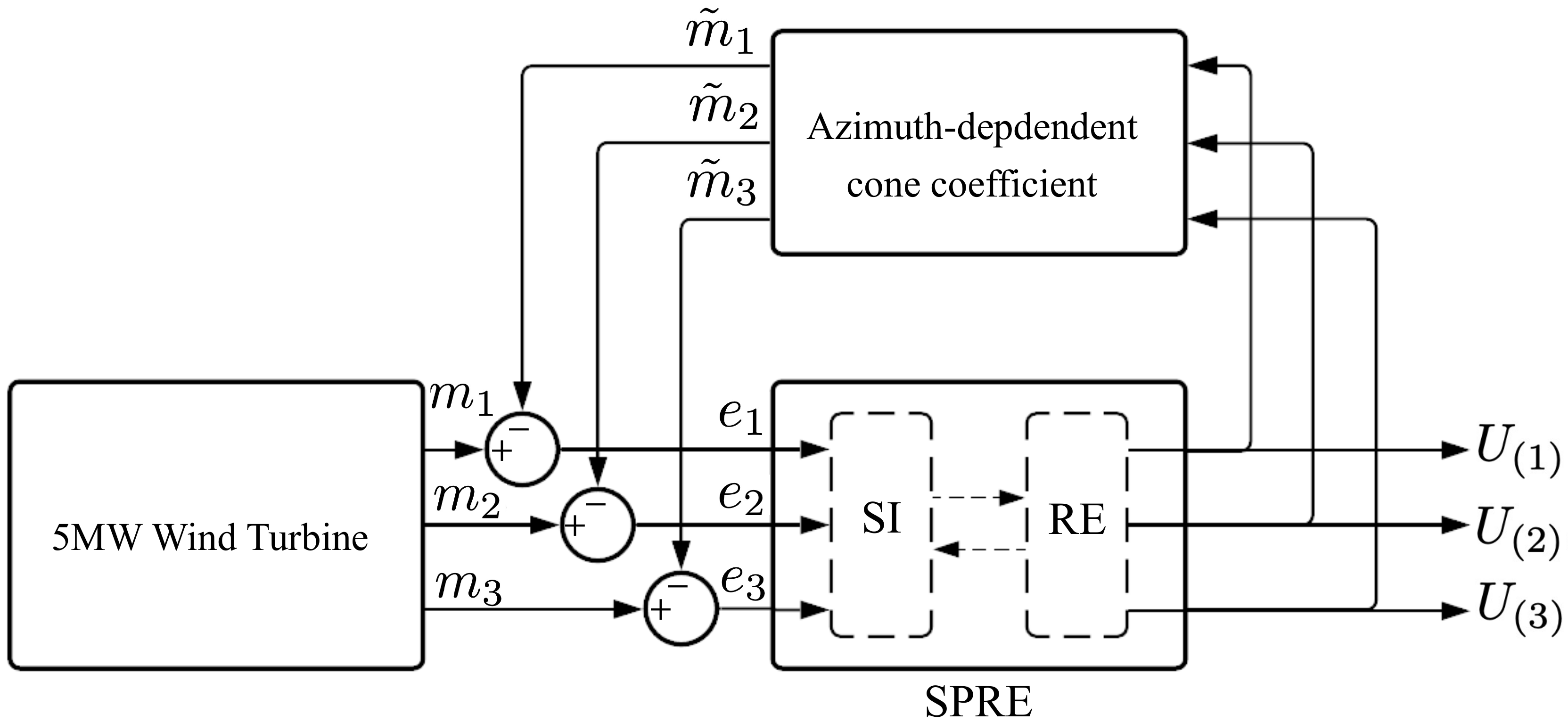}
\caption{Block diagram of the wind speed estimator and the 5MW three-bladed variable speed reference wind turbine.
The aero-structural dynamics of the wind turbine are simulated in FAST while other blocks are implemented in Simulink. SI: Subspace identification to derive a linear model, RE: repetitive estimator. $U_{(i)}$ denote the estimated BEWS, $e_i$ the error between the measured and predicted MOoP $m_i$, where $i=1,2,3$ is the blade index.}
\label{Pic_block} %
\end{figure}

The proposed wind speed estimator is comprised of an azimuth-dependent cone coefficient look-up table and the SPRE algorithm, as illustrated in Fig.~\ref{Pic_block}.
\begin{figure}
\centering 
\includegraphics[width=0.9\columnwidth]{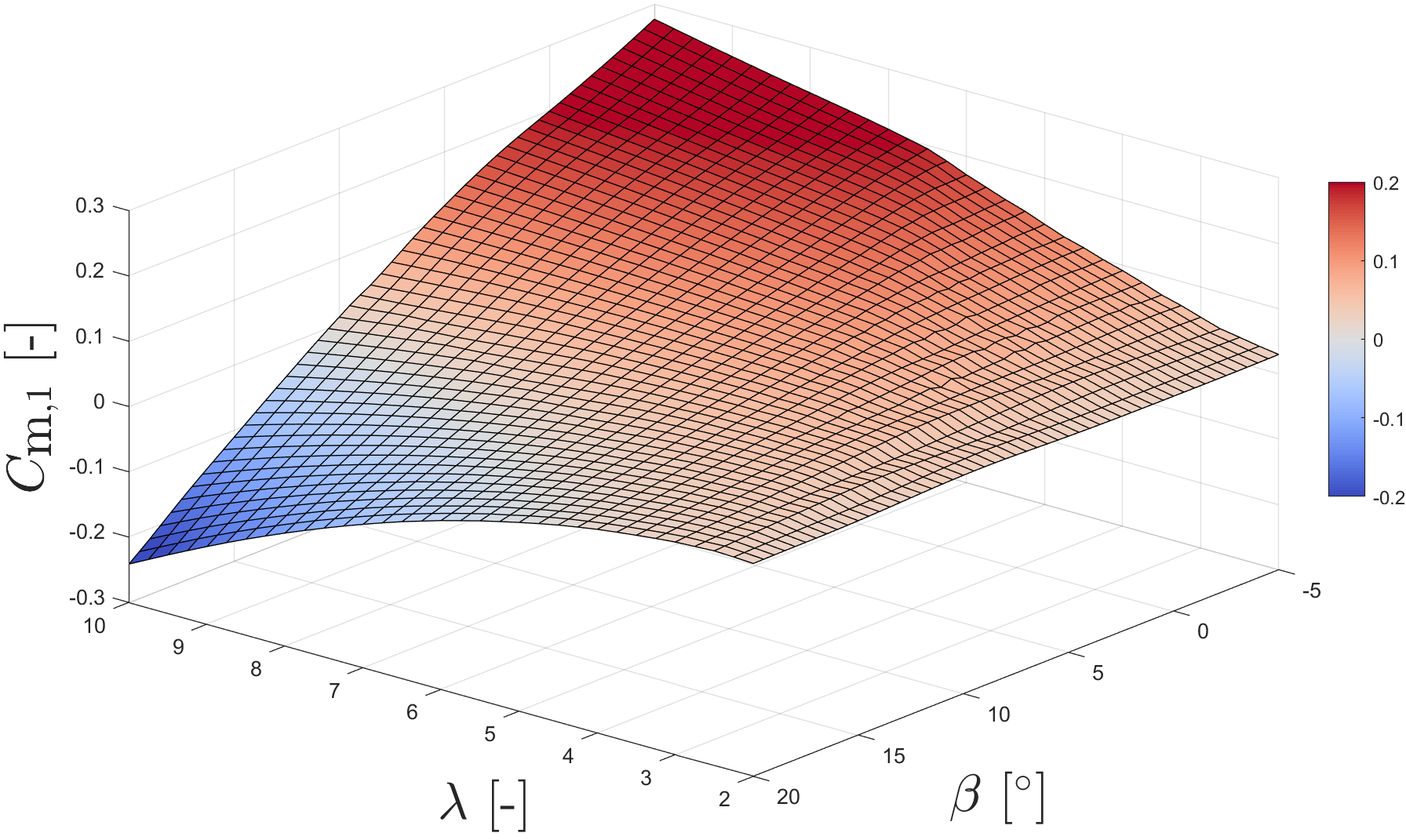}
\caption{An example of the cone coefficient of one blade at azimuth angle of $0^{\circ}$. It is computed from the steady-state wind turbine simulations where wind speed is equal to 8m/s. $\lambda$ and $\beta$ denote the tip speed ratio and the pitch angle, respectively. $C_{\text{m},1}$ is the cone coefficient of blade 1.}
\label{Cm table} %
\end{figure}
The azimuth-dependent cone coefficient is defined by extending the steady-state cone coefficient in the paper~\cite{Bottasso_2018} as
\begin{equation}
C_{m,i}(\lambda_{i}, \beta_{i}, U_{R,(i)}, \psi_i) = 
\frac{m_{i}(\lambda_{i}, \beta_{i}, U_{R,(i)}, \psi_i)}{\frac{1}{2} \rho AR U_{R,(i)}^2}
\, ,
\label{eq:cone coefficient}
\end{equation}
where $i=1,2,3$ is the blade index. $\lambda_{i}=\omega R/U_i$ denotes the blade-effective Tip Speed Ratio (TSR). 
The variable $\omega$ is the rotor speed, $R$ the rotor radius, $U_{R,(i)}$ the reference of BEWS, $\beta$ the blade pitch angle, $m_i$ the blade root out-of-plane bending moment (MOoP), $\rho$ the air density, and $A$ the rotor disk area. 
As the blades are 120$^\circ$ out of phase for a three-bladed wind turbine, the cone coefficient becomes dependent on the azimuthal position $\psi_i$ of blade $i$.
Moreover, it is also dependent on the TSR and blade pitch angle, which is similar to the widely-used power and thrust coefficients.
Due to the deformation of rotor and tower under loading, the cone coefficient is also slightly affected by the wind speed. 
In addition, the azimuth dependency is explicitly incorporated in this formulation, in order to consider the periodic components which are contributed by the gravitational force. 
The cone coefficient of one blade corresponding to TSR and pitch angle is presented in Fig.~\ref{Cm table} where the azimuth angle is fixed at zero.
It is derived from the steady-state wind turbine simulation where uniform wind speed is set to be 8m/s.
\begin{rem}\label{rem:Cm}
The azimuth-dependent cone coefficient is computed offline based on steady-state wind turbine simulations, where the uniform wind conditions without shear are considered.
Thus, only the periodic aerodynamics induced by gravity and tower shadow are included in the cone coefficient.
\end{rem}
After the cone coefficient is derived for each of the blades under all operating conditions of interests, \eqref{eq:cone coefficient} can be used to predict the MOoP $\tilde{m}_i$ based on the BEWS estimation from the SPRE approach. 
This then leads to
\begin{equation}
\tilde{m}_i = \frac{1}{2}\rho AR U_{(i)}^2C_{m,i}(\lambda_{i}, \beta_{i}, U_{(i)}, \psi_i)
\, ,
\label{eq:computation of m}
\end{equation}
where $U_{(i)}$ denotes the BEWS estimated by the SPRE approach in this paper.
The theoretical background of the proposed SPRE approach will be elaborated in Section \ref{sec:3}.

%% file: sections/3_SPRE.tex
In the proposed SPRE scheme, a discrete-time Linear Time Invariant (LTI) system along with an output predictor is formulated to describe the aerodynamics of the wind turbine.
All the matrices of the linear representation are identified recursively through an online subspace identification technique.
Based on the identified model, a repetitive estimation law is then synthesised to estimate the BEWS by solving a receding horizon optimal problem.

\subsection{Subspace Predictive Repetitive Estimator}
The wind turbine dynamics can be approximated by an LTI system affected by unknown periodic disturbances, that is, $d_k\in\mathbb{R}^m$~\cite{Houtzager_2013}.
In prediction form, it is formulated as 
\begin{equation}
\begin{cases}
\label{eq:predictor:f}
x_{k+1} \!\! &= \tilde{A}x_k+Bu_k+\tilde{E}d_k+Ly_k \\
y_k \!\! &= Cx_k+Fd_k+e_k
\end{cases} \, ,
\end{equation}
where $x_k\in\mathbb{R}^n$, $u_k\in\mathbb{R}^r$ and $y_k\in\mathbb{R}^l$ are the state, input and output vectors.
In the wind turbine model, $r = l = 3$.
The signals $u_k$ and $y_k$ include the vectors of BEWS and the error of MOoP between $\tilde{m}_i$ and $m_i$ at discrete time index $k$, respectively. 
$e_k\in\mathbb{R}^l$ is the zero-mean white innovation process or the aperiodic component of the blade loads.
$\tilde{A}\triangleq{A-LC}$ and $\tilde{E}\triangleq{E-LF}$, where $A\in\mathbb{R}^{n\times n}$, $C\in\mathbb{R}^{l\times n}$, $L\in\mathbb{R}^{n\times l}$, $E\in\mathbb{R}^{n\times m}$ and $F\in\mathbb{R}^{l\times m}$ represent the state transition, output, observer, periodic noise input and periodic noise direct feed-through matrices, respectively. $B\in\mathbb{R}^{n\times r}$ is the input matrix.
By defining a periodic difference operator $\delta$, the effect of periodic blade loads $d_k$ on the input-output system can be eliminated as
\begin{align*}
 \delta{d}_k &=d_k-d_{k-P}=0 \, ,
\end{align*} 
where $P$ is the period of the disturbance, the same as the rotation period of the rotor disk.
Similarly, $\delta{u}$, $\delta{y}$ and $\delta{e}$ can be defined as well.
Applying the $\delta$-notation to \eqref{eq:predictor:f}, this equation can be reformulated as follows, where the periodic blade load term disappears.
\begin{equation}
\begin{cases}
\delta{x}_{k+1} \!\! &= \tilde{A}\delta{x}_k+B\delta{u}_k+L\delta{y}_k \\
\delta{y}_k \!\! &= C\delta{x}_k+\delta{e}_k \, .
\end{cases} 
\label{eq:predictor2} \, .
\end{equation}
Then, a stacked vector $\delta{U}^{(p)}_{k}$ for a past time window with a length of $p$ is defined as
\begin{equation}
\delta{U}^{(p)}_{k}=
\left[ \begin{array}{c}
 u_k-u_{k-P}\\
u_{k+1}-u_{k-P+1} \\
\vdots \\
u_{k+p-1}-u_{k-P+p-1}
\end{array} 
\right ]\, .
\label{eq:stacked u}
\end{equation}
The vector $\delta{Y}^{(p)}_{k}$ can be introduced in a similar way, where 
$p$ needs to be selected large enough to ensure $\tilde{A}^{j}\approx0$ $\forall{j}\geq{p}$ \cite{Chiuso_2007}. 
With this in mind, the future state vector $\delta{x}_{k+p}$ is approximated according to $\delta{U}^{(p)}_{k}$ and $\delta{Y}^{(p)}_{k}$,
\begin{equation}
\delta{x}_{k+p} \approx
\left[ \begin{array}{cc}
K^{(p)}_u & K^{(p)}_y 
\end{array} 
\right ]
\left[ \begin{array}{c}
\delta{U}^{(p)}_{k} \\
\delta{Y}^{(p)}_{k} \\
\end{array} 
\right ] \, ,
\label{eq:lifted2}
\end{equation}
in which $K^{(p)}_u$ and $K^{(p)}_y$ are:
\begin{align*}
&\ K^{(p)}_u=
\left[ \begin{array}{cccc}
\tilde{A}^{p-1}B & \tilde{A}^{p-2}B & \cdots & B 
\end{array} 
\right ]\,,\\ 
&\ K^{(p)}_y=
\left[ \begin{array}{cccc}
\tilde{A}^{p-1}L & \tilde{A}^{p-2}L & \cdots & L
\end{array} 
\right ]\, .
\label{eq:K}
\end{align*}
By substituting this equation into \eqref{eq:predictor2}, the approximation of $\delta{y}_{k+p}$ is derived as
\begin{equation}
\delta{y}_{k+p} \approx
\underbrace{
\left[ \begin{array}{cc}
CK^{(p)}_u & CK^{(p)}_y 
\end{array} 
\right ]}_{\Xi}
\left[ \begin{array}{c}
\delta{U}^{(p)}_{k} \\
\delta{Y}^{(p)}_{k} \\
\end{array} 
\right ]
+\delta{e_{k+p}} \, .
\label{eq:lifted3}
\end{equation}
From \eqref{eq:lifted3}, it is clear that the so-called Markov matrix $\Xi$, reflects all the crucial knowledge on the wind turbine dynamics. 
It completely depends on the input vector $u$ and output vector $y$. 
Based on this, the subspace identification essentially aims to find an online solution of the following Recursive Least-Squares (RLS) optimization problem \cite{vanderVeen_2013}
\begin{equation}
\hat{\Xi}_k=\text{arg}\min_{\hat{\Xi}_k}\sum_{i=-\infty}^{k}\left \| \delta{y}_i-\gamma\hat{\Xi}_k
\left[ \begin{array}{c}
\delta{U}^{(p)}_{i-p} \\
\delta{Y}^{(p)}_{i-p} \\
\end{array} 
\right ]
\right \| ^2_2 \, ,
\label{eq:Markov parameters2}
\end{equation}
where $\gamma$ is a forgetting factor ($0\ll\gamma\leq{1}$), which aims at reducing the effect of past data, and therefore adapt to the varying system dynamics online.
In this paper, a value close to 1, \emph{i.e.,} $\gamma=0.9999$, is chosen for the identification process.
To derive a unique solution to this RLS optimization problem, a filtered pseudo-random binary signal is superimposed on the top of the input vector $u$.
Subsequently, the RLS optimization \eqref{eq:Markov parameters2} is implemented with a QR algorithm \cite{Sayed_1998} in an online recursive manner to obtain $\hat{\Xi}_{k}$.
The estimates of $\hat{\Xi}_k$ are then used in the receding horizon optimization algorithm to formulate a repetitive estimation law for the BEWS estimation.

In the repetitive estimation formulation, the wind speed is predicted over the entire rotation period. 
As $P\geq{p}$ and usually $P$ is much larger than $p$, the output equation can be lifted over the period $P$ as
\begin{equation}
\delta{Y}^{(P)}_{k+p}=\tilde{\Gamma}^{(P)}\delta{x}_{k+p}+
\left[ \begin{array}{cc}
\tilde{H}^{(P)} & \tilde{G}^{(P)} \\
\end{array} 
\right ]
\left[ \begin{array}{c}
\delta{U}^{(P)}_{k+p} \\
\delta{Y}^{(P)}_{k+p} \\
\end{array} 
\right ]
\, .
\label{eq:lift:y}
\end{equation}
$\tilde{H}^{(P)}$ is the Toeplitz matrix, which is defined as,
\begin{equation}
\tilde{H}^{(P)}=
\left[ \begin{array}{cccc}
0 & 0 & 0 & \cdots  \\
CB & 0 & 0 & \cdots  \\
C\tilde{A}B & CB & 0 & \cdots \\
\vdots & \vdots & \ddots & \vdots \\
C\tilde{A}^{p-1}B & C\tilde{A}^{p-2}B & C\tilde{A}^{p-3}B & \cdots \\
0 & C\tilde{A}^{p-1}B & C\tilde{A}^{p-2}B & \cdots \\
0 & 0 & C\tilde{A}^{p-1}B & \cdots \\
\vdots & \vdots & \ddots & \ddots \\
\end{array} 
\right ]
\, .
\label{eq:H}
\end{equation}
By replacing $B$ with $L$, $\tilde{G}^{(P)}$ can be defined as well.
The extended observability matrix $\tilde{\Gamma}^{(P)}$ is given by,
\begin{equation}
\tilde{\Gamma}^{(P)}=
\left[ \begin{array}{c}
C \\
C\tilde{A} \\
C\tilde{A}^2 \\
\vdots \\
C\tilde{A}^p \\
0 \\
\vdots \\
0
\end{array} 
\right ]
\, .
\label{eq:Gamma}
\end{equation}
In order to penalise the output $Y^{(P)}_k$ in the optimization problem, \eqref{eq:lift:y} is thus expanded as
\begin{multline}
{Y}^{(P)}_{k+P}=
\left[ \begin{array}{ccc}
I_{l{P}} & \Gamma^{(P)} \widehat{K^{(P)}_u} & \Gamma^{(P)} \widehat{K^{(P)}_y} \\
\end{array} 
\right ]
\left[ \begin{array}{c}
Y^{(P)}_k \\
\delta{U}^{(P)}_k \\
\delta{Y}^{(P)}_k 
\end{array} 
\right ]
\\
+\hat{H}^{(P)}\delta{U}^{(P)}_{k+P} \, ,
\label{eq:lift:y3}
\end{multline}
with the equalities of $\Gamma^{(P)}=(I-\tilde{G}^{(P)})^{-1}\tilde{\Gamma}^{(P)}$ and $\hat{H}^{(P)}=(I-\tilde{G}^{(P)})^{-1}\tilde{H}^{(P)}$.

In order to reduce the dimension of the state-space representation used in the receding horizon optimization framework, a B-spline basis function projection \cite{Robert_1974}, \emph{i.e.,} $\phi(\psi_i)$, is applied to \eqref{eq:lift:y3}.
The number of the B--splines is denoted as $N_b$, while the degree of the uniform knots is $N_k$ in the implementation.
Based on the basis function, the estimated wind speed can be synthesised by taking a linear combination of the B--spline basis function as
\begin{equation}
U^{(P)}_k=\phi(\psi_i) \cdot \theta_j
\, ,
\label{eq:control input}
\end{equation}
where $j=0,1,2,\cdots$ is the rotation index of the rotor. Coefficients $\theta\in\mathbb{R}^{N_b}$, which implies the weights of different splines, are updated at each $P$.
$\theta$ can be interpreted as the contribution of different splines to the wind speed estimation.
On the other hand, the output can be transformed onto the subspace that defined by the B-spline basis function, as
\begin{equation}
\bar{Y}_k=\phi^{+}(\psi_i)Y^{(P)}_k
\, ,
\label{eq:control output}
\end{equation}
in which the symbol $+$ is the Moore-Penrose pseudo-inverse. 
Based on the basis function projection, \eqref{eq:lift:y3} can be formulated into a lower dimensional state-space representation as
\begin{multline}
\underbrace{
\left[ \begin{array}{c}
\bar{Y}_{j+1}\\
\delta{\theta}_{j+1} \\
\delta{\bar{Y}}_{j+1} \\
\end{array} 
\right ]}_{\bar{\mathcal{K}}_{j+1}}
=
\underbrace{
\left[ \begin{array}{ccc}
I_{l{N_b}} & \phi^{+}\Gamma^{(P)} \widehat{K^{(P)}_u}\phi & \phi^{+}\Gamma^{(P)} \widehat{K^{(P)}_y}\phi \\
0_{l{N_b}} & 0_{r{N_b}} & 0_{l{N_b}} \\
0_{l{N_b}} & \phi^{+}\Gamma^{(P)} \widehat{K^{(P)}_u}\phi & \phi^{+}\Gamma^{(P)} \widehat{K^{(P)}_y}\phi 
\end{array} 
\right]}_{\bar{A}_j}
\\
\underbrace{
\left[ \begin{array}{c}
\bar{Y}_j \\
\delta{\theta}_j \\
\delta{Y}_j
\end{array} 
\right]}_{\bar{\mathcal{K}}_j}
+
\underbrace{
\left[ \begin{array}{c}
\phi^{+}\hat{H}^{(P)}\phi \\
I_{r{N_b}} \\
\phi^{+}\hat{H}^{(P)}\phi
\end{array} 
\right]}_{\hat{B}_j}
\delta{\theta}_{j+1}
\, .
\label{eq:state-space form_lower}
\end{multline}
Based on this, the objectives of the receding horizon optimization can be formulated in the following cost function with a sequence of future predicted coefficients $\mathbf{U} \triangleq [\delta\theta^T_{j+1}, \cdots, \delta\theta^T_{j+N_u} ]\in \mathbb{R}^{N_b \times N_u}$:
\begin{multline}
J(\bar{\mathcal{K}},\mathbf{U})=\sum_{i=0}^{N_p} 
(\bar{\mathcal{K}}_{j+i|j})^{T}Q\bar{\mathcal{K}}_{j+i|j}+ 
\\
\sum_{i=1}^{N_u} (\delta\theta_{j+i|j})^{T}R \delta\theta_{j+i|j}
\, ,
\label{eq:cost function}
\end{multline}
with the goal function of the optimization as 
\begin{equation}
V(\bar{K_j}) = \min_{\mathbf{U}} J(\bar{\mathcal{K}}_{j},\mathbf{U})
\, ,
\label{eq:MPC optimization problem}
\end{equation}
where $Q$ and $R$ are the positive-definite weighting matrices, while $N_p$ and $N_u$ are the prediction and estimation horizons, respectively.

$\mathbf{U}$ is computed by the receding horizon optimization process over the prediction horizon at each $j$. 
Only the first element $\delta \theta_{j+1}^T$ is actually selected for wind speed estimation while the remaining elements are discarded. 
$\theta_j^T$ is then synthesised according to the relation $\delta \theta_{j+1} = \theta_{j+1} - \theta_{j}$.

As a result, the predicted BEWS $U_{k+1}$ at time step $k+1$ are finally computed according to \eqref{eq:control input}.
At the next rotation count, the state $\bar{\mathcal{K}}_{j+1}$ will be updated with the online subspace identification, which is then used as an initial condition of the receding horizon optimization.
Following the philosophy of the receding horizon principle \cite{qin_2003}, the cost function in \eqref{eq:cost function} will roll ahead one step and all the procedure is repeated.

Equation~\eqref{eq:MPC optimization problem} can be solved as a standard Quadratic Program (QP) problem, by converting the optimization objectives in \eqref{eq:cost function} in the following form
\begin{equation}
J(\bar{\mathcal{K}}_j,\mathbf{U}) = X^T \mathcal{Q} X + \mathbf{U}^T \mathcal{R} \mathbf{U}
\, ,
\label{eq:QP form}
\end{equation}
where $X = [\bar{\mathcal{K}}_{j}, \bar{\mathcal{K}}_{j+1}, \cdots, \bar{\mathcal{K}}_{j+N_p}]^T$ corresponds to the vector of state predictions. $\mathcal{Q}$ and $\mathcal{R}$ are the weight matrices, which are
\begin{equation}
\mathcal{Q} = \text{diag}(Q, \cdots, Q) \,\,\,\,\,\,  \mathcal{R} = \text{diag}(R, \cdots, R)
\, ,
\label{eq:QR}
\end{equation}

By introducing the following prediction matrices, 
\begin{equation}
\mathcal{A} \!=\!
\left[   
\begin{array}
{c}
I \\
\bar{A}_j \\
\vdots  \\
{\bar{A}_j}^{N_u} \\
\vdots  \\
{\bar{A}_j}^{N_p}
\end{array}
\right ]
 , \,
\mathcal{B} \!=\! 
\left[   
\begin{array}
{ccc}
0 & \cdots & 0 \\
\hat{B}_j & \cdots & 0 \\
\vdots & \ddots & \vdots \\
{\bar{A}_j}^{N_u-1}\hat{B}_j & \cdots & \hat{B}_j \\
\vdots & \vdots & \vdots \\
{\bar{A}_j}^{N_p-1}\hat{B}_j & \cdots & \sum_{i=0}^{N_p - N_u} {\bar{A}_j}^i \hat{B}_j
\end{array}
\right ] \, ,
\end{equation}
the predictive system is introduced as
\begin{equation}
X = \mathcal{A}\bar{\mathcal{K}}_j + \mathcal{B} \mathbf{U}
\, .
\label{eq:predictive system}
\end{equation}

Combining \eqref{eq:predictive system} with \eqref{eq:cost function}, the receding horizon optimization would be implemented in the QP problem,
\begin{multline}
V(\bar{\mathcal{K}}_j) = \bar{\mathcal{K}}_j^T \mathcal{Y} \bar{\mathcal{K}}_j + \min_{\mathbf{U}} \{\mathbf{U}^T H \mathbf{U} + 2\bar{\mathcal{K}}_j^T F \mathbf{U}\}
\, ,
\label{eq:QP form2}
\end{multline}
where $H=\mathcal{B}^T \mathcal{Q} \mathcal{B} + \mathcal{R}$, $F = \mathcal{A}^T \mathcal{Q} \mathcal{B}$ and $\mathcal{Y} = \mathcal{A}^T \mathcal{Q} \mathcal{A}$.
\begin{rem}\label{rem:computational complexicity}
{\color{black}
SPRE presents low computational complexity, since the B--spline basis function is effective at reducing the dimension of the LTI system.
Moreover, the receding horizon optimization is only solved once per rotation period, which dramatically alleviates the computational burden.}
\end{rem}
By solving the QP problem in \eqref{eq:QP form2}, $\theta_j$ is computed from the receding horizon optimization. BEWS is then synthesized by taking a linear combination of the B--spline basis function in \eqref{eq:control input}. 
$\tilde{m}_i$ is subsequently estimated according to \eqref{eq:cone coefficient}-\eqref{eq:computation of m}. 
After that, the error between $m_i$ and $\tilde{m}_i$ is fed back into the SPRE algorithm for the wind speed estimation over next rotation period of the rotor. 
The approximate zero error will finally lead to a successful BEWS estimation.

%% file: sections/4_simulation.tex
The effectiveness of the proposed SPRE-based wind speed estimator is demonstrated via a case study in this section.

\subsection{Model configuration}
The wind turbine model considered in this paper is based on the 5MW three-bladed variable-speed reference wind turbine~\cite{Jonkman_2009}.
The wind turbine dynamics and baseline control system are simulated via the NREL's Fatigue, Aerodynamics, Structures, and Turbulence (FAST) tool \cite{Jonkman-2005}.
In addition, two wind inflow conditions are considered to verify this approach, which are:

(1) Stepwise sheared uniform wind flow condition: the classical power--law mean wind profile model is utilized. 
The power--law exponent is specified as 0.2. 
The amplitude of the wind speed varies from 8m/s to 15m/s.

(2) Wake-rotor overlap condition: the wind turbine is impinged by a steady-state wake shed from an upstream turbine in the farm.
An ambient wind speed of 12m/s is specified in this case.
The wind field of wake is produced by using the widely-used FLORIS model~\cite{FLORIS_2020}.
The shape of the wind farm wake is determined by the Turbulence Intensity (TI) and the center-to-center distance between turbines. In this study, TI is specified as $6\%$ while the turbine distance is 3 rotor diameters (3D).

The simulation for each wind inflow condition lasts 1000s at a fixed discrete time step of 0.01s. 
For comparison, the reference values of BEWS are derived by calculating the the wind speeds measured at two thirds of the rotor radius. 
Their average value corresponds to the real rotor effective wind speed.

\subsection{Results and discussions}
\begin{figure}
\centering 
\includegraphics[width=0.9\columnwidth]{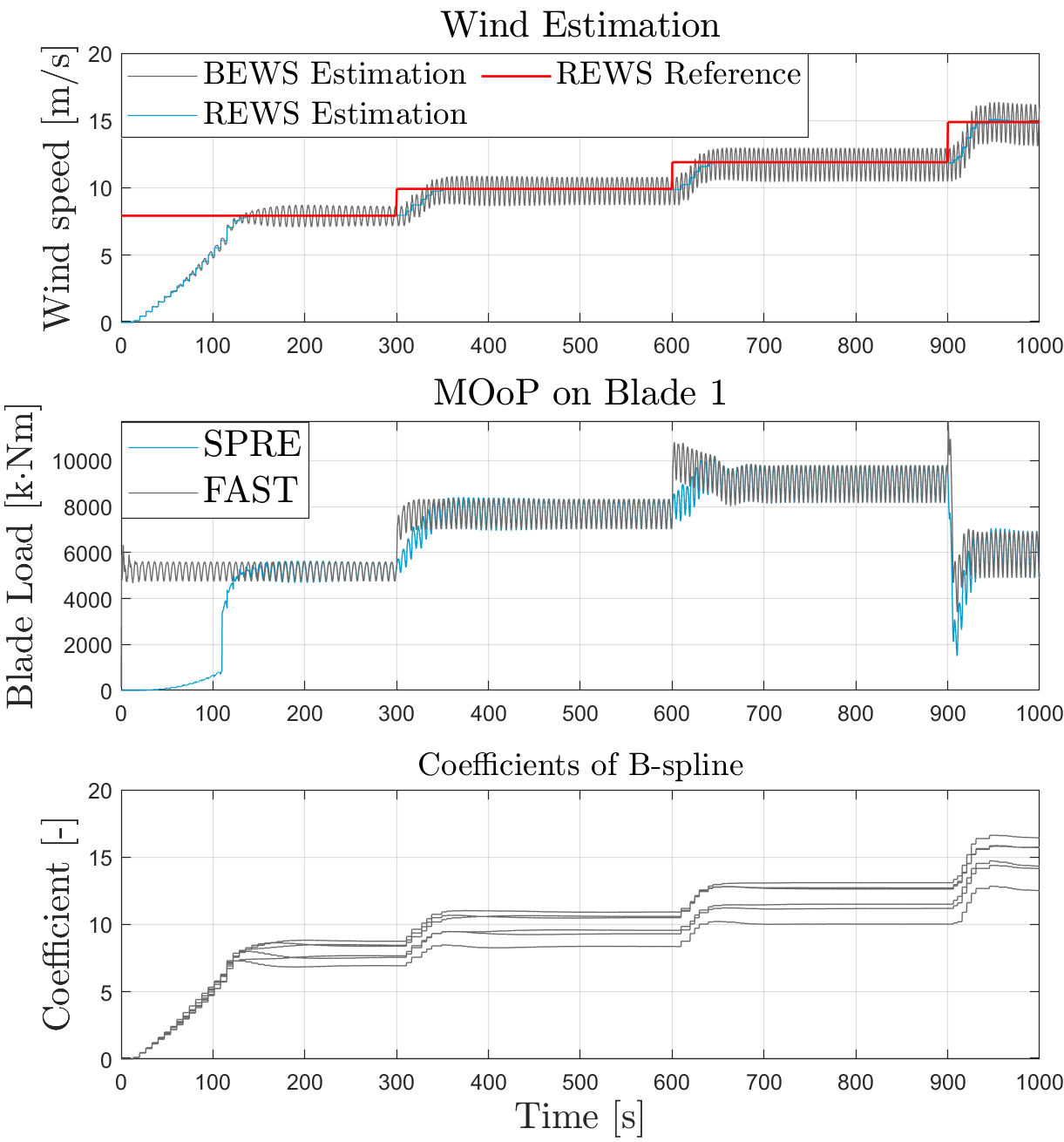}
\caption{Estimation of BEWS and REWS on blade 1 in the stepwise sheared uniform wind condition, where the hub height wind speed varies from 8m/s to 15m/s. Other blades show similar results, which are omitted for brevity.}
\label{step wind} %
\end{figure}

First of all, the performance of the proposed wind speed estimator in the first wind inflow condition is presented in Fig.\ref{step wind}.
In this case, the stepwise sheared uniform wind flow is performed.

\begin{figure}
\centering 
\includegraphics[width=0.9\columnwidth]{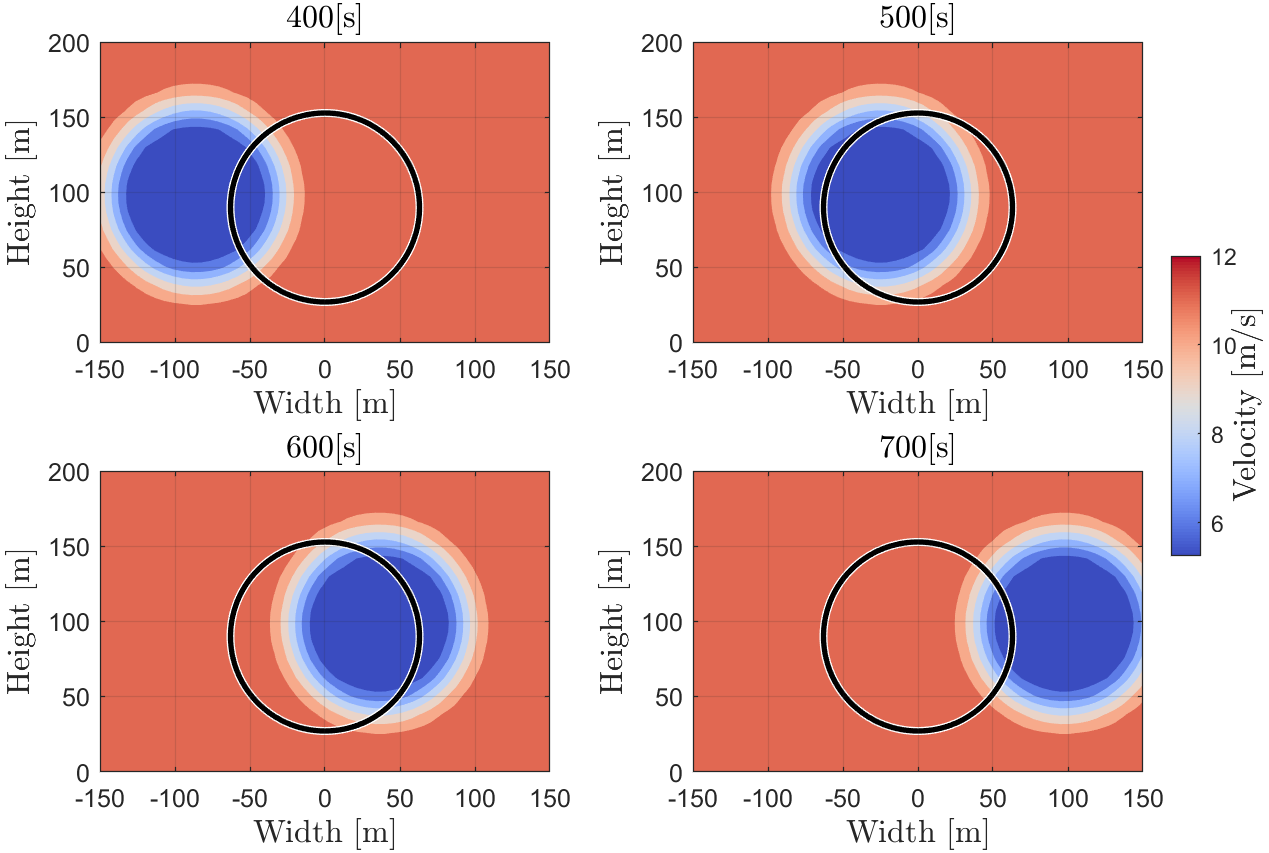}
\caption{Vertical slice of the wind inflow condition, where the rotor is impinged and overlapped by a wake. Red regions imply high wind velocity, which is undisturbed by the upstream turbine, while the blue regions indicate a velocity deficit due to the upstream turbine. The direction of the velocity vector is out-of-plane.}
\label{wake overlapping} %
\end{figure}
The performance of the BEWS estimation can be identified according to the MOoP comparisons. 
In general, the MOoP on blade 1 is effectively approximated during each step of the wind speed.
Although some deviations appear at the transition between different steps due to the abrupt increase of the wind speed, the predicted MOoP finally shows similar values as FAST simulations.
This actually suggests the successful estimation of the BEWS.
The BEWS shows significant periodic behaviors, which is induced by the wind shear over the rotational rotor disk. 
The REWS can be synthesized by calculating the average of BEWS from SPRE.
It is clear that the estimated REWS is consistent with the reference value, which implies the effective wind speed estimation.




\begin{figure}
\centering 
\includegraphics[width=0.9\columnwidth]{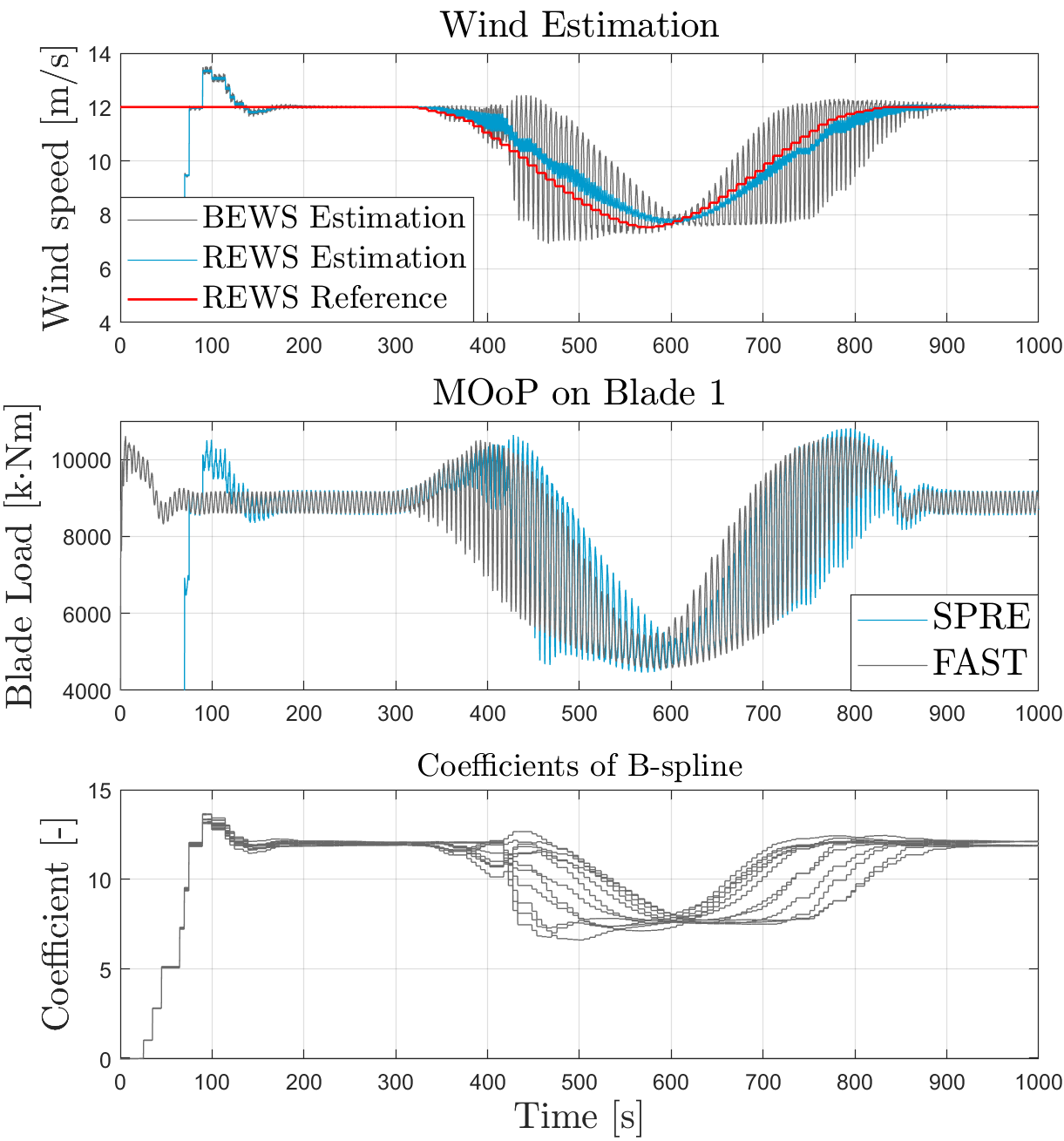}
\caption{Estimation of BEWS and REWS on blade 1 in the wake-rotor overlap condition, where the ambient wind speed is 12m/s. Other blades show similar results, which are omitted for brevity.}
\label{wake overlapping estimation} %
\end{figure}

In addition to the stepwise sheared uniform wind condition, another scenario considered in this paper is the wake-rotor overlap. 
In this case, the rotor of the wind turbine is impinged by a steady-state wake shed from the upstream turbine.
This hence leads to the partial and full wake-rotor overlap, as depicted in Fig.~\ref{wake overlapping}.
Due to the velocity deficit induced by the wake, the effective wind speed shows nonuniform and complicated patterns over blades.
It can be seen from Fig.~\ref{wake overlapping estimation} that the proposed SPRE approach presents acceptable estimates of BEWS and REWS in the wake-rotor overlap condition.
Some estimation errors are observed at the beginning of the wake-rotor overlap (350s-450s) due to the unexpected wake impingement.
The estimated MOoP can be gradually converged into real values thanks to the receding horizon optimization, thus leading to the estimates of BEWS.
The BEWS shows large fluctuations due to the partial wake-rotor overlap.
Similarly, the estimated REWS can be obtained by calculating the average of BEWS.

\begin{figure}
\centering 
\includegraphics[width=1\columnwidth]{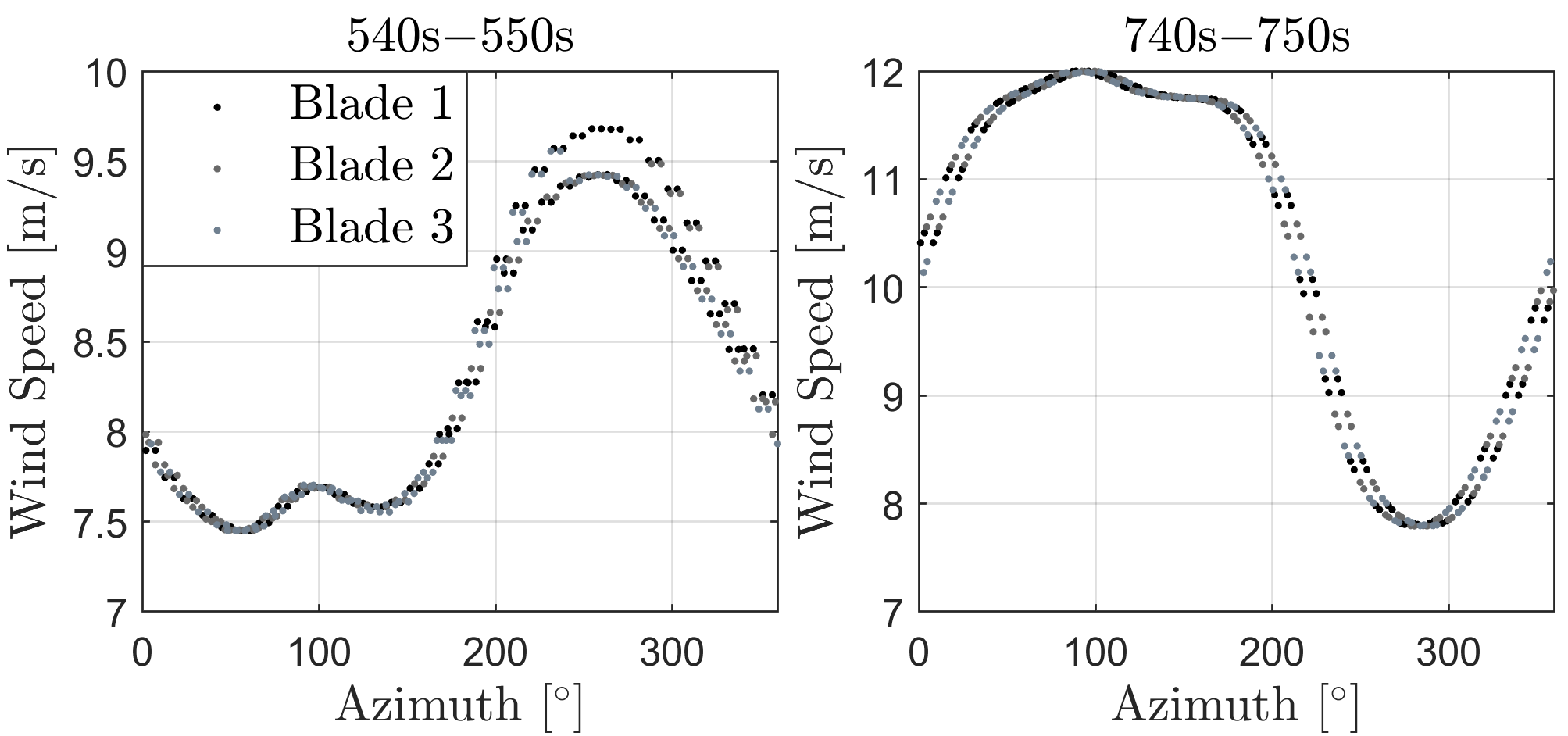}
\caption{Estimated effective wind speed over blades corresponding to the azimuth angle.}
\label{wake shape estimation} %
\end{figure}

Moreover, the estimated BEWS corresponding to the azimuth angle is demonstrated in Fig.~\ref{wake shape estimation}.
As visible, the effective wind speed over the entire rotor area shows different periodic fluctuations with azimuth angle.
The velocity deficit appears over the right sector of the rotor disk (around $100^\circ$ of azimuth angle) between 540s and 550s. Afterwards, the left sector ($290^\circ$ of azimuth angle) experiences the velocity deficit between 740 and 750s.
Combined with the vertical slice of the wind inflow condition in Fig.~\ref{wake overlapping}, this actually suggests the successful detection of the wake interference and identification of the propagation direction. 

Hence, the proposed SPRE algorithm is an effective way to estimate the BEWS in both sheared uniform wind speed and wake-rotor overlapping conditions.
By identifying the velocity deficit, it is able to detect the wake interference and perceive its propagation direction. 

%% file: sections/5_conclusions.tex
In this paper, a novel approach called Subspace Predictive Repetitive Estimator (SPRE) is proposed to estimate the Blade Effective Wind Speed (BEWS) in wind turbines.
In detail, an azimuth-dependent cone coefficient is firstly formulated to describe the nonlinear mapping between the out-of-plane blade root bending moment and effective wind speed over each blade.
Then the SPRE approach is developed to estimate the effective wind speed over blades according to the error between predicted and measured blade loads.

Case studies show that the proposed SPRE approach is effective at estimating the BEWS in the stepwise sheared uniform wind speed condition.
This approach also presents successful BEWS estimation  in a more complicated scenario where the wind turbine rotor is impinged and overlapped by a wake shed from an upstream turbine.  
By identifying the velocity deficit, it is capable of detecting the wake interference and identifying its propagation direction. 
Due to the successful estimation of the wind speeds over the blades, better knowledge of the wind can be obtained.
This will lead to a significant improvement of the control performance in modern wind turbines.
